
\input phyzzx.tex
\input myphyx.tex
\overfullrule0pt


\def\etal{{\it et al.}}

\def\third{{\textstyle{1 \over 3}}}

\def\msusy{M_{\rm SUSY}}
\def\bold#1{\setbox0=\hbox{$#1$}%
     \kern-.025em\copy0\kern-\wd0
     \kern.05em\copy0\kern-\wd0
     \kern-.025em\raise.0433em\box0 }
\Pubnum={VAND-TH-93-5}
\date={March 1993}
\pubtype{}
\titlepage

\vskip1cm
\title{\bf Top Quark and Charged Higgs: Linked by
Radiative Corrections${}^*$}

\author{Marco Aurelio D\'\i az }
\vskip .1in
\centerline{Department of Physics and Astronomy}
\centerline{Vanderbilt University, Nashville, TN 37235}
\vskip .2in

\centerline{\bf Abstract}
\vskip .1in

Some aspects of the phenomenology of the top quark and the
charged Higgs are studied, emphasizing relations due to
radiative corrections. Production mechanisms at the SSC
are mentioned, and decay modes are analyzed including radiative
corrections to the charged Higgs mass and couplings. The constraints
due to the decay $b\rightarrow s\gamma$ are discussed. Calculations
are performed in the context of the Minimal Supersymmetric
Model.

\vskip 2cm
* Talk given at the SSC Physics Symposium, University
of Wisconsin, Madison, 29-31 March 1993.

\vfill
\endpage

\voffset=-0.2cm

\noindent{\bf Introduction}

It is well known that the phenomenology of the top quark and
the charged Higgs are intimately related. If the charged Higgs
is light enough, an important production mechanism for the charged
Higgs would be the decay of the top quark $t\rightarrow bH^+$. And
at the same time, this decay mode would have an impact on the top
quark width.

\REF\neutral{M.S. Berger, {\it Phys. Rev. D} {\bf 41}, 225 (1990);
H.E. Haber and R. Hempfling,
\sl Phys. Rev. Lett. {\bf 66}, \rm 1815 (1991);
Y. Okada, M. Yamaguchi and T. Yanagida,
\sl Prog. Theor. Phys. {\bf 85}, \rm 1 (1991);
J. Ellis, G. Ridolfi and F. Zwirner, \sl Phys. Lett.
{\bf B257}, \rm 83 (1991);
R. Barbieri, M. Frigeni, F. Caravaglios, \sl Phys. Lett. {\bf B258},
\rm 167 (1991);
Y. Okada, M. Yamaguchi and T. Yanagida,
\sl Phys. Lett. {\bf B262}, \rm 54 (1991);
J.R. Espinosa and M. Quir\' oz, {\it Phys. Lett. B} {\bf 266},
389 (1991);
A. Yamada, \sl Phys. Lett. {\bf B263}, \rm 233 (1991);
J. Ellis, G. Ridolfi and F. Zwirner, {\sl Phys.
Lett.} {\bf B262}, {\rm 477 (1991)};
R. Barbieri and M. Frigeni, {\sl Phys. Lett.} {\bf
B258}, {\rm 395 (1991)};
M. Drees and M.M. Nojiri, {\it Phys. Rev. D} {\bf 45}, 2482 (1992);
A. Brignole, {\it Phys. Lett. B} {\bf 281}, 284 (1992);
M.A. D\'\i az and H.E. Haber, {\it Phys. Rev. D} {\bf 46}, 3086
(1992).}
A richer relation becomes apparent when radiative corrections are
included in the calculations. The tree level $H^+\bar tb$ vertex
depends on the quark masses, and the large value of the lower
experimental limit for the top quark mass makes radiative corrections
to  Higgs masses and couplings unusually large. This is valid also
for the neutral Higgs\refmark\neutral.

\REF\guide{J.F. Gunion, H.E. Haber, G. Kane and S. Dawson,
{\it The Higgs Hunter's Guide} (Addison-Wesley, Reading MA, 1990).}
In the context of the Minimal Supersymmetric Model
(MSSM)\refmark\guide another source of large radiative
corrections are the scalar quarks. Their mass matrix and some of their
couplings to Higgs bosons depend on the top quark mass, and if the mass
parameters in the mixing of the squark mass matrix are comparable with
the diagonal matrix elements, then large corrections are obtained.

\FIG\fgraph{Feynman diagrams related to the production and decay of
the top quark and charged Higgs: (a) $Wg\rightarrow t\bar b$, (b)
$q\bar q\rightarrow t\bar t$, (c) and (d) $gg\rightarrow t\bar t$
are the dominant production mechanisms of the top quark when
$m_t\gsim140$ GeV; (e) and (f) are two possible decay modes of the
top quark; and (g) are the main decay modes of the charged Higgs.}

\REF\topprod{C.-P. Yuan, Report No. MSUTH-92/08, December 1992,
unpublished, and references therein.}
\REF\tWbrad{B.A. Irwin, B. Margolis and H.D. Trottier, {\it Phys.
Lett. B} {\bf 256}, 533 (1991); A. Denner and T. Sack, {\it Nucl.
Phys.} {\bf B358}, 46 (1991); G. Eilam, R.R. Mendel, R. Migneron
and A. Soni, {\it Phys. Rev. Lett.} {\bf 66}, 3105 (1991);
C.-P. Yuan and T.C. Yuan, {\it Phys. Rev. D} {\bf 44},
3603 (1991).}
Top quark production mechanism at hadron colliders are
displayed in fig.~\fgraph a to \fgraph d. They are dominant for
$m_t\gsim 140$ GeV\refmark\topprod. Diagrams
\fgraph b to \fgraph d are purely
QCD, unlike diagram \fgraph a that involves electroweak interactions
as well. In fig.~\fgraph e and \fgraph f we have
some of the top quark decays: $W^+b$ and $H^+b$ (if kinematically
allowed). Electroweak radiative corrections to the first one are small
in the context of the SM (Standard Model)\refmark\tWbrad, and
electroweak corrections to the second one will be analyzed in the
following sections. In all cases the QCD corrections are implemented.

\noindent{\bf Renormalized Charged Higgs Mass}

\REF\CPRbrignole{A. Brignole, J. Ellis, G. Ridolfi and F. Zwirner,
{\it Phys. Lett. B} {\bf 271}, 123 (1991);
P.H. Chankowski, S. Pokorski and J. Rosiek,
{\it Phys. Lett. B} {\bf 274}, 191 (1992); A. Brignole, {\it Phys. Lett.
B} {\bf 277}, 313 (1992).}
\REF\diazhaberii{M.A. D\'\i az and H.E. Haber, {\sl Phys. Rev. D} {\bf
45}, \rm 4246 (1992).}
The one-loop
radiatively corrected charged Higgs mass squared is given by
\refmark{\CPRbrignole,\diazhaberii}
$$\eqalign{m_{H^{\pm}}^2&=m_W^2+m_A^2+\Delta m_{H^\pm}^2\,,\crr
\Delta m_{H^\pm}^2\equiv {\rm Re}&~\left[A_{H^+H^-}(m_W^2+m_A^2)
-A_{WW}(m_W^2)-A_{AA}(m_A^2)\right]\,.\cr}\eqn\masfordos$$
where $A$ is the CP-odd Higgs and $m_A$ is its mass.
The corrections to the squared charged Higgs mass can be approximated
in the following limit: $M_Q^2\gg m_t^2\gg m_W^2$ and
assuming $m_A^2={\cal O}(m_W^2)$, where $M_Q$ is a common
diagonal squark mass parameter. The result is
$$\eqalign{
(\Delta m_{H^{\pm}}^2)\simeq&
{{N_cg^2}\over{32\pi^2m_W^2}}\left[
{{2m_t^2m_b^2}\over{s_{\beta}^2c_{\beta}^2}}-m_W^2
\left({{m_t^2}\over{s_{\beta}^2}}+{{m_b^2}\over{c_{\beta}^2}}\right)
+{\textstyle{2\over 3}}m_W^4\right]\ln{{M_Q^2}\over{m_t^2}}\cr
-&{{N_cg^2m_t^4\mu^2}\over{64\pi^2m_W^2s_{\beta}^4M_Q^2}}\,,\hfil
\cr}\eqn\dhmi$$
where we display just the leading logarithm and the main non-leading
logarithm terms. In the last term, $\mu$ is the supersymmetric mass
parameter in the superpotential,
$\tan\beta=v_2/v_1$ is the ratio between the vacuum
expectation values of the two Higgs doublets,
and we use the notation $\sin\beta=s_{\beta}$ and similarly for other
trigonometric functions.
Complete formulas can be found in ref.~[\diazhaberii],
and are used in the analysis that follows.

\FIG\figi{Contours corresponding to (the radiatively corrected)
$m_{H^{\pm}}$ as a function of $\tan\beta$ and $m_A$.  Results
are presented for $\msusy=1$~TeV and we take the squark mixing
parameters $-\mu=A_U=A_D=\msusy$.
In (a) we fix $m_{H^{\pm}}=40$ GeV for three choices of the top
quark mass: $m_t=100$, 150 and 200~GeV. The region below the curves
is excluded by LEP data.
In (b), we plot three different contours given by $m_{H^{\pm}}=70,
100$, and 130~GeV for the top quark mass fixed at $m_t=150$~GeV.}

\REF\expmch{UA2 Collaboration, J. Alitti \etal, {\it Phys. Lett. B}
{\bf 280},
137 (1992); ALEPH Collaboration, D. Decamp \etal, {\it Phys. Rept.}
{\bf 216}, 216 (1992); UA1 Collaboration, C. Albajar \etal, {\it Phys.
Lett. B} {\bf 257}, 459 (1991); VENUS Collaboration, T. Yuzuki \etal,
{\it Phys. Lett. B} {\bf 267}, 309 (1991); DELPHI Collaboration,
P. Abreu
\etal, {\it Phys. Lett. B} {\bf 241}, 449 (1990); L3 Collaboration, B.
Adeba \etal, {\it Phys. Lett. B} {\bf 252}, 511 (1990); OPAL
Collaboration,
M. Akrawy \etal, {\it Phys. Lett. B} {\bf 242}, 299 (1990).}
{}From eq.~\dhmi\ we see that when there is an appreciable mixing
in the squark mass matrix, the radiative corrections are {\it large}
and {\it negative}. In this way, the non-observation of the charged
Higgs at LEP\refmark\expmch\
imposes constraints on the supersymmetric parameters, and can be
represented as an exclusion zone in the $\tan\beta-m_A$ plane.
In fig.~\figi\ we plot contours of constant charged Higgs
mass in the $\tan\beta-m_A$ plane. In fig.~\figi a, the contours
are for a fixed value of $m_{H^{\pm}}=40$ GeV, and the zone below
the curves is experimentally excluded. This excluded region is bigger
when the top quark mass is larger. The excluded region will grow if
the lower bound on the charged Higgs mass is increased, or if it is
discovered. This is represented in fig.~\figi b, where we see that
for low values of $\tan\beta$, radiative corrections are negative.
In this way,
the excluded region has its boundary at large values of the CP-odd
Higgs mass.
On the contrary, for large values of $\tan\beta$ radiative corrections
are positive, because terms proportional to $m_b$ not displayed in
eq.~\dhmi\ became dominant, and the boundary of the excluded region
moves to small values of $m_A$.

\noindent{\bf Renormalized ${\bold{H^+d\bar u}}$ Vertex}

\REF\diaz{M.A. D\'\i az, Report No. VAND-TH-93-1, January 1993,
to be published in {\it Phys. Rev. D};
M.A. D\'\i az, Report No. VAND-TH-92-16, to be published
in the proceedings of the 7th Meeting of the Division of Particles
and Fields of the APS, Batavia, IL, 10-14 Nov. 1992.}
The renormalized $H^+d\bar u$ vertex is\refmark\diaz:
$$\Gamma^{H^+d\bar u}(p^2)=i\lambda^-_{H^+d\bar u}(1-\gamma_5)f^-(p^2)
+i\lambda^+_{H^+d\bar u}(1+\gamma_5)f^+(p^2),\eqn\ecux$$
where the $\lambda^{\pm}$ couplings are defined by:
$$\lambda_{H^+d\bar u}^-={{gm_u}\over{2\sqrt{2}m_Wt_{\beta}}},
\qquad \lambda_{H^+d\bar u}^+={{gm_dt_{\beta}}\over{2\sqrt{2}
m_W}},\eqn\ecuvii$$
and the $f^{\pm}$ factors are complicated functions given in the
first ref. of [\diaz].

It is useful to obtain an asymptotic formula for $f^{\pm}$ by evaluating
the exact one-loop results in the same limit we used in the
previous section: $M_Q^2\gg m_t^2\gg m_W^2$ and
assuming $m_A^2={\cal O}(m_W^2)$. The result for $f^+$ is
$$f^+(m^2_{H^{\pm}})\approx 1+{{N_cg^2m_t^2}\over{128\pi^2
t_{\beta}^2m_W^2}}\eqn\fplus$$
and for $f^-$ we have
$$f^-(m_{H^{\pm}}^2)\approx\ 1+{{N_cg^2m_t^2}\over{64\pi^2m_W^2}}
\biggl[-{{c_W^2}\over{s_W^2}}+\third(r_3^2-r_1^2)-{1\over{
2t_{\beta}^2}}-\third r_1r_3{{c_{2\beta}}\over{s_{\beta}c_{\beta}}}
\biggr]\eqn\fminus$$
where we display only terms proportional to the square of the top
quark mass. The factors $r_i$ are ratios of squark mass parameters
introduced in ref. [\diazhaberii] and given by
$$r_1={{\mu+A_U/t_{\beta}}\over{M_Q}},\qquad\qquad
r_3={{\mu/t_{\beta}-A_U}\over{M_Q}}.\eqn\rdef$$

\FIG\figii{Radiatively corrected factor (a) $f^+$ and (b) $f^-$ as a
function of $\tan\beta$ at $p^2=m_{H^{\pm}}^2$.
The $f^+$ ($f^-$) factor renormalizes the term proportional to the
down(up)-type fermion mass in the $H^+d\bar u$ vertex.}

\REF\BargerHP{V. Barger, J.L. Hewett and R.J.N. Phillips,
{\it Phys. Rev. D} {\bf 41}, 3421 (1990).}
Factors $f^{\pm}$ are displayed as a function of $\tan\beta$ in
Fig.~\figii. (The perturbative region is defined by
$\tan\beta\ge m_t/$(600 GeV)\refmark\BargerHP).
Radiative corrections to $f^+$ (Fig. \figii a) are small, {\it i.e.}
$f^+$ is close to one. The two curves with small
$m_A$ are stopped at low values of
$\tan\beta$ where the radiatively corrected charged
Higgs mass reaches the lower bound of 41.7 GeV.
Unlike $f^+$, radiative corrections to $f^-$ (Fig. \figii b) can be
very important
when there is a substantial mixing in the squark mass matrix and the
parameter $\tan\beta$ is large.

\noindent{\bf Charged Higgs Fermionic Decay}

\REF\PomarolQCD{C.S. Li and R.J. Oakes, {\it Phys. Rev. D} {\bf 43},
855 (1991); A. M\'endez and A. Pomarol, {\it Phys. Lett. B}
{\bf 252}, 461 (1990).}
The dominant decay modes of a light charged Higgs are represented
in Fig.~\fgraph g (assuming that the decay mode $H^+\rightarrow W^+h$
is not kinematically allowed), and the one-loop corrected
decay rate into a quark pair is\refmark\diaz:
$$\Gamma(H^+\longrightarrow c\bar s)={{N_cg^2}\over{32\pi m_W^2}}
m_{H^{\pm}}\bigl[(f^-)^2\tilde m_c^2\cot^2\beta+(f^+)^2\tilde m_s^2
\tan^2\beta\bigr]\eqn\ecuxix$$
and to a pair of leptons:
$$\Gamma(H^+\longrightarrow \nu_{\tau}\tau^+)={{g^2}\over
{32\pi m_W^2}}m_{H^{\pm}}(f^+)^2m_{\tau}^2\tan^2\beta\eqn\ecuxx$$
Here we implement QCD corrections to the hadronic width in
eq.~\ecuxix\ by introducing running masses\refmark\PomarolQCD
$$\tilde m_q(p^2)=m_q\left[{{\ln(4m_q^2/\Lambda^2_{QCD})}\over{
\ln(p^2/\Lambda^2_{QCD})}}\right]^{12\over{33-2N_f}}\eqn\runmass$$
where $\Lambda_{QCD}$ is the usual QCD parameter that we take equal to
150 MeV, and $p^2$ is in this case the charged Higgs mass squared.
As an initial condition we take the ``threshold condition'' $\tilde
m_q(2m_q)=m_q$\refmark\PomarolQCD.

\FIG\figiii{Influence of the one-loop electroweak
radiative corrections to the charged
Higgs decay rate into a pair of fermions as a function of $\tan\beta$.
We contrast the result including only QCD corrections (dashed line) and
including also the electroweak corrections (solid line), for
the decay product $c \bar s$, and the tree level result (dashed line)
and the one-loop corrected (solid line) for $\tau^+\nu_{\tau}$.}

The decay rates are plotted in Fig.~\figiii. In the case
of the leptonic decay the radiative corrections are dominated by
the factor $f^+$; therefore they are small except for $\tan\beta<1$ and
$\tan\beta\gg 1$. In the case of the decay to two quarks, radiative
corrections are dominated by $f^+$ for $\tan^2\beta\gsim m_c/m_s$ and
by $f^-$ for $\tan^2\beta\lsim m_c/m_s$. However, the large corrections
to $f^-$ at large values of $\tan\beta$ are suppressed in the partial
width
$\Gamma(H^+\longrightarrow c\bar s)$ due to the factor $\cot^2\beta$.

\noindent{\bf Constraints due to ${\bold{b\rightarrow s\gamma}}$}

\REF\expbsf{M. Battle \etal, CLEO Collaboration, to appear in the
{\it Proceedings of the joint Lepton-Photon and Europhysics
International Conference on High Energy Physics}, Geneva, Switzerland,
August 1991.}
\REF\hewettBBP{J.L. Hewett, {\it Phys. Rev. Lett.} {\bf 70}, 1045
(1993); V. Barger, M.S. Berger and R.J.N. Phillips, {\it Phys. Rev.
Lett.} {\bf 70}, 1368 (1993).}
\REF\InamiL{T. Inami and C.S. Lim, {\it Prog. Theor. Phys.} {\bf 65},
297 (1981).}
\REF\TwoHDM{T.G. Rizzo, {\it Phys. Rev. D} {\bf 38}, 820 (1988); B.
Grinstein and M.B. Wise, {\it Phys. Lett. B} {\bf 201}, 274 (1988);
W.-S. Hou and R.S. Willey, {\it Phys. Lett. B} {\bf 202}, 591 (1988);
T.D. Nguyen and G.C. Joshi, {\it Phys. Rev. D} {\bf 37}, 3220 (1988);
C.Q. Geng and J.N. Ng, {\it Phys. Rev. D} {\bf 38}, 2857 (1988);
D. Ciuchini, {\it Mod. Phys. Lett. A} {\bf 4}, 1945 (1989);
B. Grinstein, R. Springer and M. Wise, {\it Nucl. Phys.} {\bf B339},
269 (1990); T. Hayashi, M. Matsuda and M. Tanimoto, report No.
AUE-01-93, unpublished.}
\REF\BBMR{S. Bertolini, F. Borzumati, A. Masiero and G. Ridolfi,
{\it Nucl. Phys.} {\bf B353}, 591 (1991).}
\REF\susy{S. Bertolini, F. Borzumati and A. Masiero, {\it Nucl. Phys.}
{\bf B294}, 321 (1987); F.M. Borzumati, Report No. PRINT-93-0025
(Hamburg), to be published in the Proceedings of XV Meeting on
Elementary Particle Physics, Kazimierz, Poland, May 1992.}
\REF\qcd{S. Bertolini, F. Borzumati and A. Masiero, {\it Phys. Rev.
Lett.} {\bf 59}, 180 (1987); N.G. Deshpande, P. Lo, J. Trampetic,
G. Eilam and P. Singer, {\it Phys. Rev. Lett.} {\bf 59}, 183 (1987);
B. Grinstein, R. Springer and M.B. Wise, {\it Phys. Lett. B}
{\bf 202}, 138 (1988); P. Cho and B. Grinstein, {\it Nucl. Phys.}
{\bf B365}, 279 (1991).}
\REF\diazbtosf{M.A. D\'\i az, Report No. VAND-TH-93-2, January 1993,
to be published in {\it Phys. Lett. B}.}
The upper bound \refmark\expbsf
on the branching fraction of the inclusive decay $b\rightarrow
s\gamma$, given by $B(b\rightarrow s\gamma)<8.4\times 10^{-4}$,
sets powerful constraints on the charged Higgs sector
\refmark\hewettBBP. This decay is forbidden at tree level,
but induced at the one-loop level
\refmark{\InamiL-\susy}.
QCD corrections to the process $b\rightarrow s\gamma$ have been
found to be large\refmark\qcd, as well as the influence of the
electroweak corrections to the charged Higgs mass and to the
charged-Higgs-fermion-fermion vertex, in the context of the MSSM
\refmark\diazbtosf.

Following ref.~[\BBMR], at the leading order, the SM and the charged
Higgs contributions to the amplitude of the process $b\rightarrow
s\gamma$ are proportional to the operator $m_b\epsilon_{\mu}\bar s
i\sigma^{\mu\nu}q_{\nu}b_R$, where $\sigma^{\mu\nu}=(i/2)[\gamma
^{\mu},\gamma^{\nu}]$ and $q$ and $\epsilon$ are
the four-momentum and polarization of the outgoing
photon. The corrected coefficient $A_{tot}$ is\refmark\diazbtosf
$$A_{tot}=A_{SM}+(f^+f^-)A_{H^{\pm}}^1+(f^-)^2\cot^2\beta
A_{H^{\pm}}^2\eqn\corratot$$
where we evaluate the functions $f^{\pm}(p^2)$ at the typical energy
scale of this decay process $p^2=m_b^2$. Functions $A^1_{H^{\pm}}$
and $A^2_{H^{\pm}}$ are available in ref. [\BBMR].

\FIG\figiv{Contours of constant branching fraction
$B(b\rightarrow s\gamma)$
in the $\tan\beta-m_A$ plane. We display three values for the
branching fraction:
$8.4\times10^{-4}$, $7.0\times10^{-4}$ and $5.5\times10^{-4}$.
We plot the branching fraction including corrections to the charged
Higgs mass and to the charged-Higgs-fermion-fermion vertex
(solid line),
only corrections to the charged Higgs mass (dashed line),
and the pure one-loop answer with tree level mass and coupling
(dot-dashed line).
All the supersymmetry breaking mass parameters in the squark sector
are taken to be equal to $M_{SUSY}$.}

We plot in Fig.~\figiv\ contours of
constant value of the branching fraction $B(b\rightarrow s\gamma)$:
the pure one-loop answer, the one-loop plus
corrections only to the charged Higgs mass, and
corrections to both the charged Higgs mass and the
charged-Higgs-fermion-fermion vertex are displayed.
The excluded zone lies below the solid line corresponding to
$B=8.4\times10^{-4}$. We see that the inclusion
of electroweak radiative corrections to the Higgs sector has an
important effect when there is a substantial mixing in the squark
mass matrix. In particular, we cannot rule out a light CP-odd Higgs.

\noindent{\bf Top Quark Decay to a Charged Higgs}

\REF\treetdecay{V. Barger and R.J.N. Phillips, {\it Phys. Rev. D}
{\bf 41}, 884 (1990); R.M. Godbole and D.P. Roy, {\it Phys. Rev.
D} {\bf 43}, 3640 (1991).}
\REF\QCDtbH{C.S. Li and T.C. Yuan, {\it Phys. Rev. D} {\bf 42}, 3088
(1990); M. Dress and D.P. Roy, {\it Phys. Lett. B} {\bf 269}, 155
(1991); C.S. Li, Y.S. Wei and J.M. Yang, {\it Phys. Lett. B} {\bf 285},
137 (1992); J. Liu and Y.-P. Yao, {\it Phys. Rev. D} {\bf 46}, 5196
(1992); A. Czarnecki and S. Davidson, Report No. Alberta-THY-34-92,
December 1992, unpublished.}
\REF\susyQCDtbH{C.S. Li, J.M. Yang and B.Q. Hu, Report No.
CQU-TH-92-16, October 1992, unpublished.}
\REF\THDMtbH{C.S. Li, B.Q. Hu and J.M. Yang, Report No. CQU-TH-92-14,
unpublished.}
\REF\MPomarol{A. M\'endez and A. Pomarol, {\it Phys. Lett. B}
{\bf 265}, 177 (1991).}
If $m_t>m_{H^{\pm}}+m_b$, the decay mode
of the top quark $t\rightarrow H^+b$
becomes important and in some cases dominates over the decay mode
$t\rightarrow W^+b$\refmark\treetdecay. QCD corrections have been
studied\refmark{\QCDtbH,\susyQCDtbH}, as well as electroweak
corrections in a non-supersymmetric model\refmark\THDMtbH.
The corresponding branching fractions satisfy
$$\eqalign{
{{B(t\rightarrow bH^+)}\over{B(t\rightarrow bW^+)}}=&
{{\lambda^{1/2}(m_t^2,m_b^2,m_{H^{\pm}}^2)}\over{
\lambda^{1/2}(m_t^2,m_b^2,m_W^2)}}\times\cr&
{{(m_t^2f^{-2}/t_{\beta}^2+m_b^2f^{+2}t_{\beta}^2)
(m_t^2+m_b^2-m_{H^{\pm}}^2)+4m_t^2m_b^2f^+f^-}\over{
m_W^2(m_t^2+m_b^2-2m_W^2)+(m_t^2-m_b^2)^2}}\cr}
\eqn\rbron$$
where we have introduced the electroweak corrections to the charged
Higgs vertex in the MSSM defined by the functions $f^{\pm}$.
The function $\lambda$ is defined by $\lambda(a,b,c)=
a^2+b^2+c^2-2ab-2ac-2bc$.
If we use the correction factors $f^{\pm}$ valid for
the charged Higgs vertex to a pair of light fermions, we are
neglecting some terms that are reported to be small in the context of
a non-supersymmetric two-Higgs-doublets model\refmark\MPomarol.

\FIG\figv{Ratio between the branching fractions of the top quark
decay modes $t\rightarrow bH^+$ and $t\rightarrow bW^+$ as a function
of $\tan\beta$, for two different values of the CP-odd charged
Higgs mass: $m_A=40$ GeV and $m_A=100$ GeV. We include electroweak
and QCD radiative corrections (solid line) and compare with the case
in which only QCD corrections are implemented (dashed line).}

In Fig.~\figv\ we plot the ratio between the branching fractions
$B(t\rightarrow bH^+)$ and $B(t\rightarrow bW^+)$ as a function of
$\tan\beta$. We contrast the effect of the electroweak radiative
corrections by including only QCD effects (dashed line) and both
QCD and electroweak effects (solid line). In order to implement QCD
effects, we replace the bottom quark mass by a running mass as
indicated by eq.~\runmass. The electroweak corrections are important
at $\tan\beta\lsim 1$ and $\tan\beta\gsim 10$.

\noindent{\bf Conclusions}

In the MSSM leading logarithm radiative corrections to the charged
Higgs mass are proportional to $m_t^2$. Nevertheless, when there is a
substantial mixing in the squark mass matrix, radiative corrections
are large and negative due to a non-leading term proportional to
$m_t^4$. This effect and the experimental lower bound on $m_{H^{\pm}}$
rules out a small region in the parameter space. This region
will grow as the bound for $m_{H^{\pm}}$
increases (or if it is discovered).

The squark mass mixing effect is also very important in the radiative
corrections to the charged Higgs coupling to a pair of light fermions.
Large corrections are found if there is an appreciable squark mixing,
specially if $\tan\beta<1$ or $\tan\beta\gg 1$, but these
corrections to the coupling are suppressed in the
decay rate of the charged Higgs into a pair of light fermions, due to
powers of $\tan\beta$ in the denominator.
On the contrary, the squark mixing effect is very important in the
calculation of the one-loop induced decay $b\rightarrow s\gamma$
when the corrected charged Higgs mass and coupling are used. The upper
bound on the branching fraction $B(b\rightarrow s\gamma)$ imposes
constraints on the parameter space that are altered by the two-loop
corrections described above. In particular, a light CP-odd Higgs is not
excluded at large $\tan\beta$.

The effect of the electroweak radiative corrections on the top
quark decay $t\rightarrow H^+b$ is studied also. Calculating the
ratio $B(t\rightarrow H^+b)/B(t\rightarrow W^+b)$, where the $W$ boson
and the charged Higgs are produced on-shell, we appreciate the
importance of the radiative corrections at large and small values of
$\tan\beta$.

{\bf \noindent Acknowledgements }
\vglue 0.4cm
I am thankful to Howard E. Haber for his contribution to part of
the work presented here. I have benefited also with
discussions with Howard Baer, Ralf Hempfling, Alex Pomarol
and Thomas Weiler. This work was supported in part by the U.S.
Department of Energy, grant No. DE-FG05-85ER40226.
\vskip 0.4cm

\endpage
\refout
\figout
\end